# An Energy Efficient Multichannel MAC Protocol for Cognitive Radio Ad Hoc Networks

S. M. Kamruzzaman

School of Electronics and Information Engineering
Hankuk University of Foreign Studies, Yongin-si, Kyonggi-do, 449-791 Korea
*smzaman@hufs.ac.kr*

**Abstract**: This paper presents a TDMA based energy efficient cognitive radio multichannel medium access control (MAC) protocol called ECR-MAC for wireless Ad Hoc Networks. ECR-MAC requires only a single half-duplex radio transceiver on each node that integrates the spectrum sensing at physical (PHY) layer and the packet scheduling at MAC layer. In addition to explicit frequency negotiation which is adopted by conventional multichannel MAC protocols, ECR-MAC introduces lightweight explicit time negotiation. This two-dimensional negotiation enables ECR-MAC to exploit the advantage of both multiple channels and TDMA, and achieve aggressive power savings by allowing nodes that are not involved in communication to go into doze mode. The IEEE 802.11 standard allows for the use of multiple channels available at the PHY layer, but its MAC protocol is designed only for a single channel. A single channel MAC protocol does not work well in a multichannel environment, because of the multichannel hidden terminal problem. The proposed energy efficient ECR-MAC protocol allows SUs to identify and use the unused frequency spectrum in a way that constrains the level of interference to the primary users (PUs). Extensive simulation results show that our proposed ECR-MAC protocol successfully exploits multiple channels and significantly improves network performance by using the licensed spectrum band opportunistically and protects QoS provisioning over cognitive radio ad hoc networks.

## 1. Introduction

Present wireless networks are based on a static or fixed spectrum assignment policy that is regulated by government agencies, has let to a quasi-scarcity of the spectrum. Traditionally, spectrum segments are licensed on a long term basis in particular geographic regions. Only small segments of unlicensed spectrum remain available. Cognitive radio (CR) [1] technology has been proposed as a promising solution to share the scarce spectrum resources in an opportunistic way while avoiding disruptions to the legacy devices of wireless networks i.e TV broadcast stations and wireless microphone. The CR user also called secondary user (SU) is allowed to use only locally unused spectrum so that it does not cause any interferences or collisions to the incumbent or primary users (PUs). Recent spectrum measurements [2] show that fixed spectrum policy is becoming unsuitable for today's wireless communications. As the frequency spectrum becomes exhausted [3], CR is becoming a hot research topic in the wireless communications arena.

Cognitive radio networks (CRNs) refer to networks where nodes are equipped with a spectrum agile radio which has the capabilities of sensing the available spectrum band, reconfiguring radio frequency, switching to the selected frequency band and use it efficiently without interference to PUs [4] [5]. CR Ad hoc networks (CRANs) are emerging, infrastructure less multi-hop CRNs. The CR users (nodes) can communicate with each other through ad hoc connection.

The throughput of multi-hop wireless networks can be significantly improved by multichannel communications compared with single channel communication, as transmission can be processed on different channels simultaneously while avoiding collisions and interference in wireless ad hoc networks [6] [7]. We consider a multichannel CRN, in which every node is equipped with single network interface card (NIC) and can be tuned to one of the available channels. A pair of NICs can communicate with each other if they are on the same channel and are within the transmission range of each other.

Although the basic idea of CR is simple, the efficient design of CRNs imposes the new challenges that are not present in the traditional wireless networks [8]–[10]. Specifically, identifying the time-varying channel availability imposes a number of nontrivial design problems to the MAC layer. One of the most difficult, but important, design problems is how the SUs decide when and which channel they should tune to in order to transmit/receive the SUs' packets without interference to the PUs. This problem becomes even more challenging in wireless ad hoc networks where there are no centralized controllers, such as base stations or access points.

As CRNs need to use several channels in parallel to fully utilize the spectrum opportunities, the MAC layer should accordingly be designed. Multichannel MAC protocols have clear advantages over single channel MAC protocols: They offer reduced interference among users, increased network throughput due to simultaneous transmissions on different channels, and a reduction of the number of CRs affected by the return of a licensed user [11]. By exploiting multiple channels, we can achieve a higher network throughput than using single channel, because multiple transmissions can take place without interfering. Designing a MAC protocol that exploits multiple channels is not an easy task, due to the fact that each of current IEEE 802.11 device is equipped with one half-duplex transceiver. The transceiver is capable of switching channels dynamically, but it can only transmit or listen on one channel at a time. Thus, when a node is listening on a particular channel, it cannot hear communication taking place on a different channel. Due to this, a new type of hidden terminal problem occurs in this multichannel environment, which we refer to as multichannel hidden terminal problem. So a single channel MAC protocol (such as IEEE 802.11 DCF) does not work well in a multichannel environment where nodes may dynamically switch channels.

To amend the aforementioned problems of the existing schemes, in this paper, we propose multichannel ECR-MAC protocol which enables nodes to dynamically negotiate channels such that multiple communications can take place in the same region simultaneously, each in different channel. The network we consider is an ad hoc network that does not rely on infrastructure, so there is no central authority to perform channel management. To coexist with the licensed PUs in an ad hoc based multichannel CR environment and to achieve a higher throughput, one of the important issues is to utilize multiple channels on the licensed band efficiently while causing little interference to PUs. The main idea is to divide time in to fixed-time intervals using beacons, and have a small window at the start of each interval to indicate traffic and negotiate channels and time slots for use during the interval. A similar approach is used in IEEE 802.11 power saving mechanism (PSM) [12], explained in section 3.2. The proposed scheme can eliminates contention between nodes, decomposes contending traffics over different channels and timeslots based on actual traffic demand. As a result, the proposed scheme leads to significant increases in network throughput and decreases the end-to-end delay in an energy efficient way.

## 2. Related Work

The underutilization of spectrum under the current static spectrum management policy has stimulated a flurry of existing research activities in searching CR MAC protocols. Recently, several attempts were made to develop MAC protocols for CRNs [13]-[21]. One of the key challenges to enabling CR communications is how to perform opportunistic medium access control (MAC) while limiting the interference imposed on PUs. The IEEE 802.22 working group is in the process of standardizing a centralized MAC protocol that enables spectrum reuse by CR users (a.k.a SUs) operating on the TV broadcast bands [22]. In [17]-[19], centralized protocols were proposed for coordinating spectrum access. For an ad hoc CRN without centralized control, it is desirable to have a distributed MAC protocol that allows every CR user to individually access the spectrum.

A number of multichannel contention-based MAC protocols were previously proposed in the context of CRNs [13]-[16]. The CRN MAC protocol in [13] jointly optimizes the multichannel power/rate assignment, assuming a given power mask on CR transmissions. How to determine an appropriate power mask remains an open issue. Distance and traffic-aware channel assignment (DDMAC) in cognitive radio networks is proposed in [14]. It is a spectrum sharing protocol for CRNs that attempts to maximize the CRN throughput through a novel probabilistic channel assignment algorithm that exploits the dependence between the signal's attenuation model and the transmission distance while considering the prevailing traffic and interference conditions.

A bandwidth sharing approach to improve licensed spectrum utilization (AS-MAC) is presented in [15] is a spectrum sharing protocol for CRNs that coexists with a GSM network. CR users select channels based on the CRN's control exchanges and GSM broadcast information. Explicit coordination with the PUs is required. In [21], the authors developed a spectrum-aware MAC protocol for CRNs (CMAC). CMAC enables opportunistic access and sharing of the available white spaces in the TV spectrum by adaptively allocating the spectrum among contending users.

A distributed cognitive radio MAC (DCR-MAC) protocol is proposed in [23] for wireless ad hoc networks that provides for the detection and protection of incumbent systems around the communication pair. DCR-MAC operates over a separate common control channel and multiple data channels; hence, it is able to deal with dynamics of resource availability effectively in cognitive networks. A simple and efficient sensing information exchange mechanism between neighbor nodes with little overhead is proposed. A cognitive MAC protocol for multichannel wireless networks (C-MAC) is proposed in [24], which operates over multiple channels, and hence is able to effectively deal with the dynamics of resource availability due to PUs and mitigate the effects of distributed quiet periods utilized for PU signal detection. In C-MAC, each channel is logically divided into recurring superframes which, in turn, include a slotted beaconing period (BP) where nodes exchange information and negotiate channel usage. Each node transmits a beacon in a designated beacon slot during the BP, which helps in dealing with hidden nodes, medium reservations, and mobility.

CR based multichannel MAC protocols for wireless ad hoc networks (CRM-MAC) is proposed in [25], which integrate the spectrum sensing and packet scheduling. In their protocols each SU is equipped with two transceivers. One of the transceivers operates on a dedicated control channel, while the other is used as a CR that can periodically sense and dynamically utilize the identified unused channels. CR-enabled multichannel MAC (CREAM-MAC) protocol is proposed in [26], which integrates the spectrum sensing at physical layer and packet scheduling at MAC layer, over the wireless networks. In the proposed CREAM-MAC protocol, each SU is equipped with a CR-enabled transceiver and multiple channel sensors. The proposed CREAM-MAC enables the SUs to best utilize the unused frequency spectrum while avoiding the collisions among SUs and between SUs and PUs.

Distributed CR MAC (COMAC) protocol is presented in [27] that enable unlicensed users to dynamically utilize the spectrum while limiting the interference on PUs. The main novelty of COMAC lies in not assuming a predefined SU-to-PU power mask and not requiring active coordination with PUs. COMAC provides a statistical performance guarantee for PUs by limiting the fraction of the time during which the PUs' reception is negatively affected by CR transmissions. To provide such a guarantee, COMAC develop probabilistic models for the PU-to-PU and the PU-to-SU interference under a Rayleigh fading channel model. From these models, they derive closed-form expressions for the mean and variance of interference.

A distributed multichannel MAC protocol for multi-hop CRNs (MMAC-CR) is proposed in [28] that look at CR-enabled networks with distributed control. In addition to the spectrum scarcity, energy is rapidly becoming one of the major bottlenecks of wireless operations and has to be considered as a key design criterion. They present an energy-efficient distributed multichannel MAC protocol for CR networks. Decentralized cognitive MAC (DC-MAC) for dynamic spectrum access is presented in [29] is a cross-layer distributed scheme for spectrum allocation/sensing. It provides an optimization framework based on partially observable Markov decision processes, with no insights into protocol design, implementation, and performance.

A CR MAC protocol using statistical channel allocation for wireless ad hoc networks (SCA-MAC) is presented in [30]. SCA-MAC is a CSMA/CA based protocol, which exploits statistics of spectrum usage for decision making on channel access. For each transmission, the sender negotiates with the receiver on transmission parameters through the control channel. Synchronized MAC protocol for multi-hop CRNs (SYN-MAC) is proposed in [31], where the use of common control channel (CCC) is avoided. The scheme is applicable in heterogeneous environments where channels have different bandwidths and frequencies of operation.

## 3. Multichannel Hidden Terminal Problem

Normally, when a node is neither transmitting nor receiving, it listens to the control channel. When node A wants to transmit a packet to node B, A and B exchange RTS and CTS messages to reserve the channel as in IEEE 802.11 DCF [12]. RTS and CTS messages are sent on the control channel. When sending an RTS, node A includes a list of channels it is willing to use. Upon receiving the RTS, node B selects a channel and includes the selected channel in the CTS. After that, node A and B switch their channels to the agreed data channel and exchange the DATA and ACK packets.

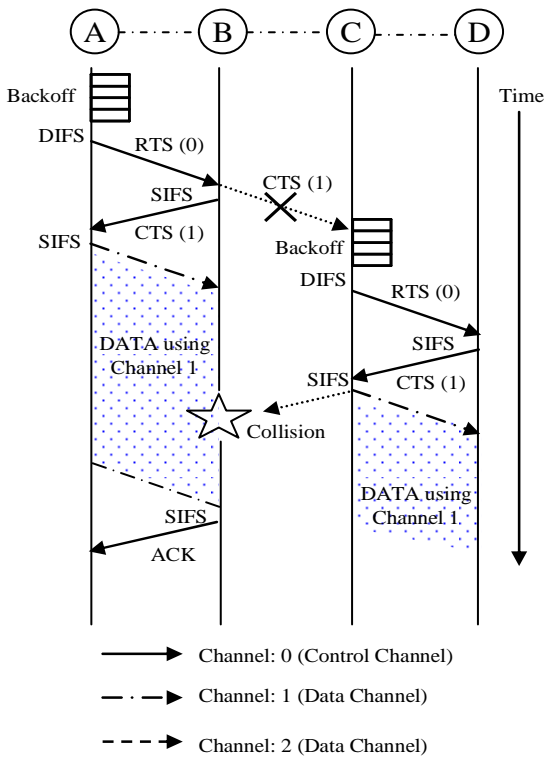

**Figure 1.** Multichannel hidden terminal problem

Now consider the scenario in figure 1. Node A has a packet for B, so A sends an RTS on channel 0 which is the control channel. B selects channel 1 for data communication and sends CTS back to A. The RTS and CTS messages should reserve channel 1 within the transmission ranges of A and B; so that no collision will occur. However, when node B sent the CTS to A, node C was busy receiving on another channel, so it did not hear the CTS. Not knowing that B is receiving on channel 1, C might initiate a communication with D, and end up selecting channel 1 for communication. This will result in collision at node B. The above problem occurs due to the fact that nodes may listen to different channels, which makes it difficult to use virtual carrier sensing to avoid the hidden terminal problem. If there was only one channel that every node listens to, C would have heard the CTS and thus deferred its transmission. Thus, we call the above problem the multichannel hidden terminal problem. As presented in the section 5, we solve this problem using synchronization, similar to IEEE 802.11 power saving mechanism (PSM) [12].

## 4. System Model

We consider a multi-hop CRANs composed of a set of CR users, each of which is equipped with a single half-duplex CR transceiver. We assume CR users are stationary or moving very slowly. In our CRN, PUs are also assumed to be stationary and they coexist with the CR users. Each PU operates with an ON–OFF switching cycle that is unknown to the CRN. Consider the spectrum consisting of $C$ non-overlapping channels, each with bandwidth $B_c$ ($c = 1, 2, ..., C$). These $C$ channels are licensed to PUs. CR can dynamically access any one channel to deliver its packets. Considering the fact that the spectrum opportunity is changing frequently with time and locations, we assume that CR users exchange control information in a dedicated channel which is always available. This dedicated channel may be owned by the CR service provider [32].

We assume that each transceiver always transmits at a fixed transmission power and hence, their transmission range $R_c$ and interference range $I_c$, which is typically 2 to 3 times of transmission range [33], are fixed for a particular channel $c$. We use a communication graph $G(V, E)$, to model the network where each node $v \in V$ corresponds to a CR user in the network and $E$ is the set of communication links each connecting a pair of nodes. There is a link $l = (u, v) \in E$ between nodes $u$ and $v$, if two nodes are in the transmission range and there is an available channel $c \in C_u \cap C_v$. Where $C_u$ and $C_v$ represent list of available channels at node $u$ and $v$ respectively. A communication link $l = (u, v)$ denotes that $u$ can transmit directly to $v$ if there are no other interfering transmissions. Due to the broadcast nature of the wireless links, transmission along a link may interfere with other link transmissions when transmitted on the same channel but links on different channels do not interfere.

An interference model defines which set of links can be active simultaneously without interfering. We model the impact of interference by using the well known protocol model proposed in [34]. We say a transmission from a transmitter in node $u$ can be successfully received by a receiver in node $v$ on a certain channel at some time instant, if

$$\frac{G_{uv}P_{uv}}{N_o + \sum_{(x,y)\in\tau\setminus\{(u,v)\}}G_{xv}P_{xy}} \geq \beta. \qquad (1)$$

In Inequality (1), $\tau$ stands for the set of concurrent transmissions; $P_{uv}$ is the power level set at the transmitter of node $u$ for transmission $(u, v)$; $G_{uv}$ is the channel gain for node pair $(u, v)$ depending on path loss, channel fading and shadowing; $\beta$ is a given threshold determined by some QoS requirements such as bit error rate (BER); $N_0$ is the thermal noise power at the receiver of node $v$ which is usually a small constant. The left hand side of this inequality is normally called the signal to interference and noise ratio (SINR) at the receiver of node $v$. Note that the SINR constraint (Inequality (1)) is satisfied at each receiver implies that the half-duplexing, unicasting and collision-free constraints are satisfied at each receiver.

A transmission on channel $c$ through link $l$ is successful if all interferes in the neighbourhood of both nodes $u$ and $v$ are silent on channel $c$ for the duration of the transmission. Two wireless links $(u, v)$ and $(x, y)$ interfere with other if they work on the same channel and any of the following is true: $v = x$, $u = y$, $v \in Nb(x)$, or $u \in Nb(y)$. Where $Nb(v)$ represents the set of neighbors of node $v$. If links $(u, v)$ and $(x, y)$ are conflicting, nodes $u$ and $y$ are within two-hops of each other [35]. The interference model can be represented by a conflict graph $F$ whose vertices corresponds to the links in the communication graph, $G$. There is an edge between two vertices in $F$ if the corresponding links can not be active simultaneously. Two links sharing a common node conflict with each other, and will have an edge in between. In addition, links in close proximity will interfere with each other if they are assigned with the same channel and hence connected with edges.

## 5. ECR-MAC Design

A TDMA scheme is used in the communication window of our proposed ECR-MAC as depicted in the figure 2. The ECR-MAC scheme has some similarities with TMMAC [36]. We assume that time domain is divided into fixed length beacon intervals and each beacon interval consists of an ad hoc traffic indication messages (ATIM) window, a sensing window, and a communication widow. The ATIM window is contention-based and uses the same mechanism as in the IEEE 802.11 DCF [12]. The ATIM window is divided into the beacon and the control window. During the ATIM window, control channel is used for beaconing and to exchange control message. All of the CR users are synchronized by periodic beacon transmissions. In this MAC scheme, channel sensing is performed before data transmission to avoid possible collisions with PUs. If any chosen channel is found to be busy, the corresponding CR users will switch to the control channel and wait until the next beacon interval. Otherwise it will go for data transmission in communication window.

As mentioned earlier, the communication window is time-slotted and uses TDMA scheme. The duration of each timeslot is the time required to transmit or receive a single data packet and it depends on the data rate of PHY layer and the size of data unit. In order to minimize possible collision with transmission from PUs, the slot size is restricted for a single data packet. The duration of the timeslot is long enough to accommodate a data packet transmission, including the time need to switch the channel, transmit the data packet and the acknowledgement. According to our MAC structure, the duration of each slot is $D_{slot} = D_{data} + D_{ACK} + 2 \times D_{guard}$. The use of guard period is to accommodate the propagation delay and the transition time from $T_x$ mode to $R_x$ mode. In the communication window, nodes can send or receive packets or go to sleep mode to save power.

If a node has negotiated to send or receive a packet in the $j^{th}$ time slot, it first switches to the negotiated channel and transmits or waits for the data packet in that slot. If a receiver receives a unicast packet, the receiver sends back an ACK in the same time slot as shown in the slot structure of figure 2. Note that proposed ECR-MAC scheme does not guarantee 100% collision-free communication in the communication window, since packet collision may occur in the ATIM window which may convey incorrect information of negotiation. If a sender does not hear an ACK after it sends a unicast packet, may be because of the collision with other transmissions, the sender may perform random backoff before attempting its retransmission using free time slots. If the number of retransmissions exceeds the retry limit, the packet is dropped. It is noted that along with other channels control channel can also be used for data transmission in the communication window as shown in figure 3, if needed. If a node has not negotiated to send or receive a data packet in the $j^{th}$ time slot, the node switches to doze mode for power saving.

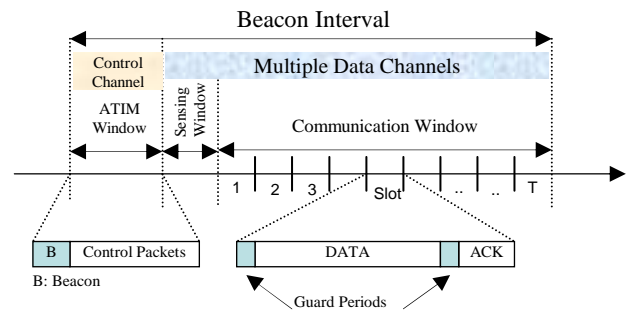

**Figure 2.** Structure of ECR-MAC protocol

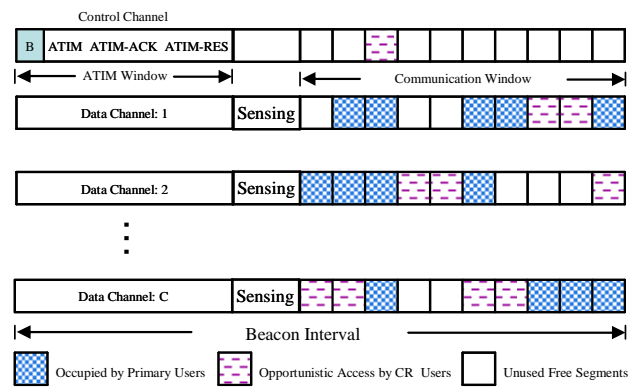

**Figure 3.** Process of channel negotiation and data exchange in ECR-MAC

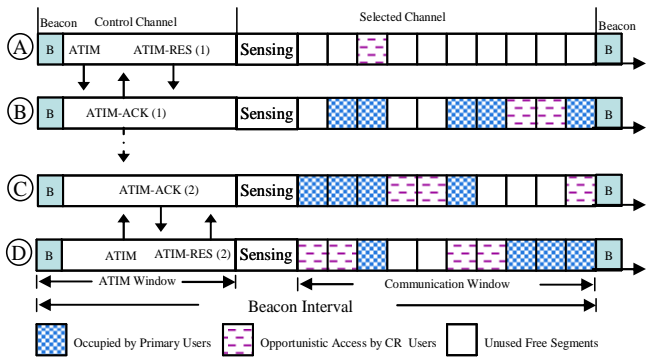

**Figure 4.** Solution of multichannel hidden terminal problem using ECR-MAC protocol

To assure collision-free communications, all neighborhood nodes of the intended receiver except the intended transmitter should remain silent on the particular channel during a given timeslot. With the help of periodic beaconing, each node is aware of (1) the identities and list of available channels within its two-hop neighbor, and (2) existing transmission schedule of communication segments of its one-hop neighbor. Based on the collected neighbor information and its own information each secondary node updates the status of its communication segments as occupied or free. Free communication segment of node *v, free_segment(v)*, is defined as the communication segments for all available channels, which are used by node *v* to communicate with adjacent nodes, and are not interfere by other transmissions. Status of the communication segments on a link is determined by finding the intersection of the status of both end nodes of the link.

For each link in the network, the communication segment assignment algorithm marks each communication segment as one of the following:

- Occupied: this segment is using by other transmissions and hence can not be used.
- Free: unassigned idle segment.
- Assigned: this segment shall be used for packet transmission on a specific link.

We define the set of common free communication segments between two nodes to be the link bandwidth. If we let $B(u, v)$ be the available bandwidth of the link between nodes *u* and *v* then $B(u, v) = free\_segment(u) \bigcap free\_segment(v)$.

Suppose that node *A* has packets for *B* and thus *A* sends an ATIM packet to *B* during the ATIM window, with *A*'s free communication segment list included in the packet. On receiving the ATIM request from *A*, *B* decides which segment(s) to use during the beacon interval, based on its free communication segments and *A*'s communication segments.

The communication segment (channel-timeslot) selection procedure is discussed in the next sub section. After selecting the channel and time slot(s), *B* sends an ATIM-ACK packet to *A*, specifying the channel and time slot(s) it has chosen. When *A* receives the ATIM-ACK packet, *A* will see if it can also select the channel-timeslot specified in the ATIM-ACK packet. If it can, it will send an ATIM-RES packet to *B*, with *A*'s selected channel-timeslot specified in the packet. If *A* cannot select the channel which *B* has chosen, it does not send an ATIM-RES packet to *B*. The process of channel-timeslot negotiation and data exchange in ECR-MAC is illustrated in figure 3. Figure 4 shows how multichannel hidden terminal problem can be solved by using our ECR-MAC protocol. During the ATIM window, *A* sends ATIM to *B* and B replies with ATIM-ACK indicating to use channel 1 and timeslot(s). This ATIM-ACK is overheard by C, so channel 1 will not be selected by *C*. When *D* sends ATIM to *C*, *C* selects channel 2 and timeslot(s). So, after the ATIM window, the two communications (between *A* and *B*, and *C* and *D*) can take place simultaneously in communication window.

### 5.1 Selection of Communication Segments

In this subsection, we present a heuristic algorithm to select communication segments for the link $l = (u, v)$. Let us consider $r_r(z)$ be the remaining data rate requirement for the session *z* of a connection request. Initially $r_r(z) = r(z)$. The basic idea of this approach is to select minimum number of free communication segments to satisfy the given rate requirement within the interference constraint. In order to maintain minimum number of communication segments in a link we will use high capacity segments. Sort all the free communication segments in the descending order of their capacities. Pick a communication segment $(c, t)$ from the sorted list and check the capacity of the chosen segment $\alpha(c, t) = B_c / |T|$ is not less than the $r_r(z)$, then it is selected. The selected segment is then removed from the free segment list and update the remaining rate requirement $r_r(z)$. To ensure the collision-free transmissions, the following conditions must be satisfied in selecting the communication segments. Let segment $(c, t)$ is trying to assign for the link $l = (u, v)$ such that:

- Timeslot *t* is not assigned to any link incident (connected) on node *u*,
- Timeslot *t* is not assigned to any outgoing link from node *v*,
- Timeslot *t* is not used on channel *c* by any link $l'$, $T_x(l') \in Nb(v)$, where $Nb(v)$ represents the set of neighbors of node *v*; and
- Timeslot *t* is not used on channel *c* by any link $l'$, $R_x(l') \in Nb(u)$.

Without confusions, $T_x(\cdot)/R_x(\cdot)$ represent both the transmitter/receiver of the given link. Note that one of the necessary constraints for collision-free communication is that no two links incident at node can be assigned same timeslot [35]. If all the above conditions are satisfied, communication segment $(c, t)$ is assigned to the link $l = (u, v)$. This procedure continues until the rate requirement is satisfied.

## 6. Performance Evaluation

The effectiveness of the proposed ECR-MAC protocol is validated through simulation. This section describes the simulation environment, performance metrics, and experimental results. The result of our approach is compared with SYN-MAC [31], SCA-MAC [30], CREAM-MAC [26], and IEEE 802.11 DCF [12]. We used network simulator-2 (NS-2) version *ns-2.33* [37] to evaluate the performance of the proposed ECR-MAC protocol. We generate 10 random topologies, and the result is the average over the 10 random topologies. The simulated network is composed of 80 static

CR nodes deployed randomly within a $1000m \times 1000m$ square region. Based on the IEEE 802.11a standard, the number of channels is set to 12 including 11 data channels and one control channel. The data channels are divided into three groups that include 3 channels in the first group and 4 channels each in last two groups. Based on the IEEE 802.11b, data rates for these groups are set to 2 Mbps, 5.5 Mbps, and 11 Mbps. Nodes can respectively transmit 1, 3, or 5 consecutive packets depending on their channel condition. The data rate for control channel is 2 Mbps.

The transmission and interference range of each CR user (node) is approximately $150m$ and $300m$ respectively. The control channel can support a transmission range of $200m$. We set initial energy as 60 joules per node. The number of timeslots in the communication window is set to 20 and the length of the ATIM window is 20ms. Channel switching delay for CR transceiver is $40\mu s$. We randomly placed 5 PUs in the region. Each of them randomly chose a channel to use, which is then considered to be unavailable for all the CR users within their coverage range, which is set to 300m. We initiate sessions between randomly selected but disjoint source-destination pairs. The two-ray-ground reflection model is used to propagation model. The maximum transmission power is set to $P_{max} = 300mW$. The thermal noise power is set to $N_0 = -90dBm$. The SINR threshold is set to $\beta = 10dB$. The channel gain, $G_{uv}$ is set to $1/d_{uv}^4$, where $d_{uv}$ is the Euclidean distance between node $u$ and node $v$. The traffic demand for each communication session is given by a random number uniformly distributed in $[0.1B_c, 0.6B_c]$, where $B_c$ is the channel capacity of channel $c$. The packet size of each flow is set to 1000 bytes (excluding the size of IP layer and MAC layer headers). Data traffic was generated using constant bit rate (CBR) traffic sources generating 4 packets/second. All traffic sessions are established at random times near the beginning of the simulation run and they stay active until the end. Simulations are run for 500 simulated seconds. The following performance metrics are used to evaluate the proposed protocol:

*Normalized Throughput:* The ratio of throughput obtains using CRN routing protocols to the throughput obtain when using IEEE 802.11 DCF on a single channel environment. The normalized throughput quantifies the performance improvement of CRN (multichannel) protocols with respect to a single channel network.

*Average End-to-End Delay:* Average latency incurred by the data packets between their generation time and their arrival time at the destinations.

*Per Packet Energy:* Per packet energy is the value of total energy consumed by the whole network divided by the total number of data packets successfully transmitted to the destinations.

In the first simulation, we measured the normalized throughput varying the number of flows shown in figure 5. The throughput of ECR-MAC is compared with other protocols including IEEE 802.11 DCF single channel network using UDP traffic. The number of simultaneous UDP flows is varied from 2 to 30. As we can see from the figure, when the number of flows increases, ECR-MAC offers significantly better performance than all other protocols especially compared with IEEE 802.11 DCF. The throughput of ECR-MAC is 7.4 times that of IEEE 802.11 DCF. When the network is overloaded, ECR-MAC achieves

8% more throughput than CREAM-MAC, 26% more than SCA-MAC, and 72% more than SYN-MAC protocol. Throughput of SYN-MAC is less because there is no CCC for conveying the control messages. As a result many connection requests are dropped resulting less throughput.

In addition, when the number of flows is large, the available channel diversity can be better exploited. Furthermore, when the number of flows is increased, ECR-MAC can significantly improve the network throughput. That's because the channel assignment algorithm can balance the channel load to different channels. So the traffic is allocated on different channels in an approximate average manner. Finally, ECR-MAC achieves higher performance because ECR-MAC eliminates inter-flow and intra-flow interference using a non-conflicting channel-timeslot assignment.

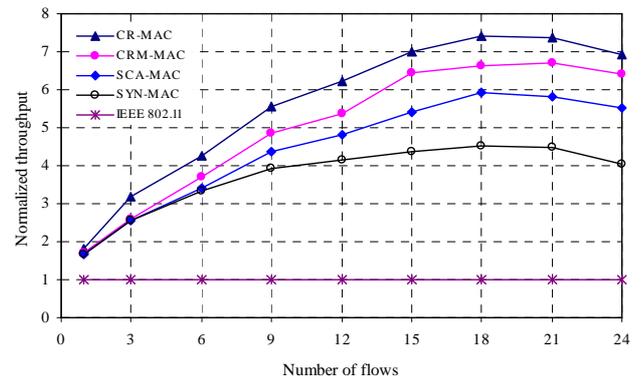

**Figure 5.** Network throughput varying number of flows

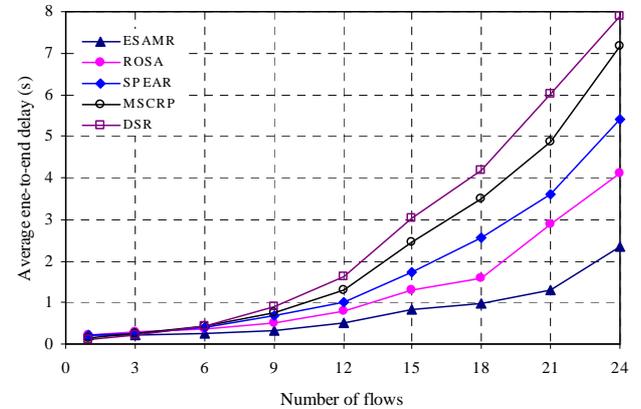

**Figure 6.** Average end-to-end delay varying number of flows

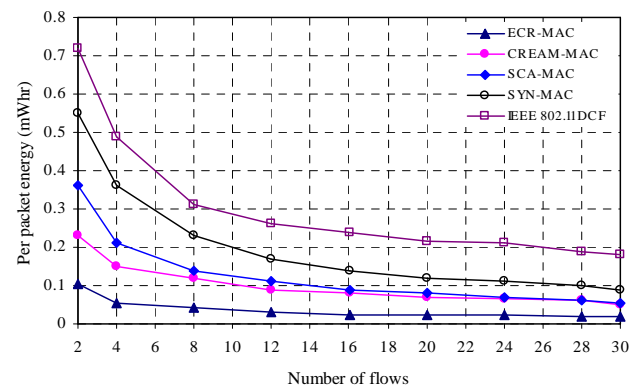

**Figure 7.** Per packet energy varying number of flows

Figure 6 shows the average end-to-end packet delay of the protocols as the network load increases. The difference between IEEE 802.11 DCF and other protocols in delay is due to the fact that with only one channel, a packet has to wait longer to use the channel when the network load is high. When comparing with other protocols ECR-MAC shows lower delay in all network scenarios. IEEE 802.11 DCF achieves better performance than other schemes when the number of flows is less. However, according to increase of number of flows, queuing delay is raised. The queuing delay makes the performance of each protocol worse. Specially, the end-to-end packet transmission delay of IEEE 802.11 is increased dramatically according to increase of flows because IEEE 802.11 uses only a single channel for every data transmission. On the other hand, the data traffic is split into multiple channels in the case of ECR-MAC. Therefore the end-to-end packet transmission delay of ECR-MAC is increased slowly according to increase of flows.

Figure 7 shows that ECR-MAC consumes much less per packet energy compared to other protocols. When the number of flows is 2, per packet energy consumption in ECR-MAC is 14% of IEEE 802.11 DCF, 19% of SYN-MAC, 28% of SCA-MAC, and 44% of that in CREAM-MAC. The energy savings in ECR-MAC becomes more significant as the number of flows increases. We conclude the following reasons for the low per packet energy consumption in ECR-MAC. Firstly, ECR-MAC allows a node to switch to doze mode in a time slot whenever it is not scheduled to transmit or receive a packet. In other protocols, due to the lack of time negotiation, a node needs to stay awake during the whole communication window when it has negotiated to transmit or receive packets. Finally, ECR-MAC achieves much higher aggregate throughput, which further reduces its per packet energy consumption.

## 7. Conclusion

In this paper, we present the ECR-MAC protocol, which is a TDMA based energy efficient multichannel MAC protocol using a single half duplex transceiver for cognitive radio ad hoc networks. ECR-MAC requires time synchronization in the network in order to avoid the multichannel hidden terminal problem and divides time into fixed beacon intervals. Nodes that have packets to transmit negotiate which channels and time slots to use for data communication with their destinations during the ATIM window. This negotiation enables ECR-MAC to exploit the advantage of both multiple channels and TDMA in an efficient way. In addition, ECR-MAC is able to support broadcast in an energy effective way. Since ECR-MAC only requires one transceiver per node, it can be implemented with hardware complexity comparable to IEEE 802.11.

Though ECR-MAC protocol has some similarities with TMMAC but it has several features that really distinguish ECR-MAC from TMMAC or any other related MAC protocols. First of all, ECR-MAC is designed for CR ad hoc networks but TMMAC was designed for traditional multichannel networks. We introduce communication segment allocation algorithm to ensure collision free communication in our protocol which is not addressed in TMMAC. In addition the broadcast procedure of ECR-MAC is also different from TMMAC. We used an efficient way to broadcast the message through the broadcast slot of ATIM window. Furthermore, our protocol addresses the QoS requirement.

Simulation results show that ECR-MAC successfully exploits multiple channels to improve network throughput and the end-to-end delay. Extensive simulations show the effectiveness of the ECR-MAC and demonstrate its capability to provide high throughput and low end-to-end delay for robust multi-hop communications in an energy efficient way that meets the QoS requirements while the other protocols could not. The analytical model of this protocol is not included in this paper. However, it is kept for future extension of this work with more simulation results.

## Acknowledgment

The author would like to thank Professor Dong Geun Jeong for his valuable time and suggestions during the completion of this research work. Author also wishes to thank the anonymous reviewers for their valuable comments to improve the quality of the paper.